\RequirePackage{fix-cm}% fix the issues with T1-encoded CModern fonts
\RequirePackage{fixltx2e} %general LaTeX2e bugfixes: should be loaded before amsthm, subfig
\documentclass[aps,pre,twocolumn,superscriptaddress,10pt,nofootinbib]{revtex4-1}

\usepackage{subfigure}
\usepackage{amsfonts, amssymb, amsmath}
\usepackage{bm}
\usepackage{graphics}
\usepackage{graphicx}
\usepackage{wasysym}
\usepackage{natbib}
\usepackage{mathtools}

\usepackage[utf8x]{inputenc}
\usepackage[T1]{fontenc}
\usepackage{paralist}

\usepackage[usenames,dvipsnames]{xcolor}

\usepackage{color}
\definecolor{darkblue}{rgb}{0,0,0.6}
\definecolor{darkred}{rgb}{0.6,0,0}

\usepackage{tikz}
\usetikzlibrary{decorations.pathreplacing}
\usepackage{hyperref}
\hypersetup{
bookmarksopen=true,
pdftitle= Activating critical exponent spectra with a slow drive, % title of the pdf
pdfauthor= S. Mathey et al., % authors of the pdf
pdftoolbar=false, % toolbar hidden
pdfstartview={FitH},		% fits the width of the page to the window
pdfmenubar=true,			%menubar shown
pdfhighlight=/O,			%effect of clicking on a link
colorlinks=true,			%couleurs sur les liens hypertextes
urlcolor=darkblue,
citecolor=darkblue,		%couleur des liens pour les citations
linkcolor=MidnightBlue,	%couleur des liens hypertextes internes
}
\usepackage[normalem]{ulem}
\usepackage{comment}
\newcommand{\bvec}[1]{{\boldsymbol{#1}}}
\newcommand{\ie}{i.e.~}
\newcommand{\iec}{i.e.,~}

\newcommand{\eg}{e.g.~}
\newcommand{\egc}{e.g.,~}

\newcommand{\Fig}[2]{Fig.~\ref{#1}\textcolor{MidnightBlue}{#2}}

\newcommand{\Sect}[1]{Sec.~\ref{#1}}
\newcommand{\App}[1]{App.~\ref{#1}}
\newcommand{\Eq}[1]{Eq.~(\ref{#1})}
\newcommand{\eq}[1]{(\ref{#1})}
\newcommand{\Eqs}[1]{Eqs.~(\ref{#1})}

\usepackage{abbrevs}
\usepackage{etoolbox} % for \robustify
\newabbrev\RG{Renormalization Group (RG)}[RG]
\newabbrev\KZ{Kibble-Zurek (KZ)}[KZ]
\robustify{\RG}
\robustify{\KZ}

\makeatletter %This fixed a but with abbrevs.
\renewcommand\maybe@space@{%
  \maybe@ictrue % <= this is new
  \expandafter   \@tfor
    \expandafter \reserved@a
    \expandafter :%
    \expandafter =%
                 \nospacelist
                 \do \t@st@ic
  \ifmaybe@ic % <= this is new
    \space
  \fi
}
\makeatother

\begin{document}

\title{Activating critical exponent spectra with a slow drive}

\author{Steven Mathey}
\email[]{smathey@thp.uni-koeln.de}
\affiliation{Institut f\"ur Theoretische Physik, Universit\"at zu K\"oln, 50937 Cologne, Germany}
\affiliation{Kavli institute for theoretical physics, Kohn Hall, University of California, Santa Barbara CA 93106-4030}

\author{Sebastian Diehl}
\affiliation{Institut f\"ur Theoretische Physik, Universit\"at zu K\"oln, 50937 Cologne, Germany}
\affiliation{Kavli institute for theoretical physics, Kohn Hall, University of California, Santa Barbara CA 93106-4030}

\date{\today}

%_____________________________________________________________________________________________________
\begin{abstract}
We uncover an aspect of the Kibble--Zurek phenomenology, according to which the spectrum of critical exponents of a classical or quantum phase transition is revealed, by driving the system slowly in directions parallel to the phase boundary. This result is obtained in a renormalization group formulation of the Kibble--Zurek scenario, and based on a connection between the breaking of adiabaticity and the exiting of the critical domain via new relevant directions induced by the slow drive. The mechanism does not require fine tuning, in the sense that scaling originating from irrelevant operators is observable in an extensive regime of drive parameters. Therefore, it should be observable in quantum simulators or dynamically tunable condensed-matter platforms.
\end{abstract}

%_____________________________________________________________________________________________________
\maketitle

\section{Introduction}

Universality near classical and quantum second-order phase transitions builds on the fact that long-wavelength fluctuations are freed by the fine tuning to the critical point, and dominate the macroscopic physics. This is quantified by only few independent observable critical exponents, universal numbers which are independent of microscopic details and determined exclusively by the system's symmetries and dimensionality. However, universality extends beyond these so-called relevant exponents to an entire spectrum of universal exponents \cite{cardy1996scaling,Pelissetto2002,tauber2014critical}. Yet, this fully universal information is not easily accessible in static experiments (see, \egc \cite{Kaul1988,Kaul1991,Kaul1994,Pelissetto2002}), or even numerics, the subleading character of the associated power laws is easily overwritten by the more dominant exponents (see, \egc \cite{Bloete1988,Li1995,Ballesteros1996,Grassberger1997,Hasenbusch2010,Kaupuzs2017,Clisby2017,Ferrenberg2018}). Exceptions to this scenario are available in conformal field theories, where a relation between the scaling dimensions of operators and the energy spectrum has been established \cite{Cardy1984,Cardy1986,Trebst2008,Laeuchli2013} as well as in transitions with dangerously irrelevant parameters, where specific exponents are accessible \cite{Leonard2015}.

\begin{figure}[t]
\begin{center}
 \includegraphics[width=\columnwidth]{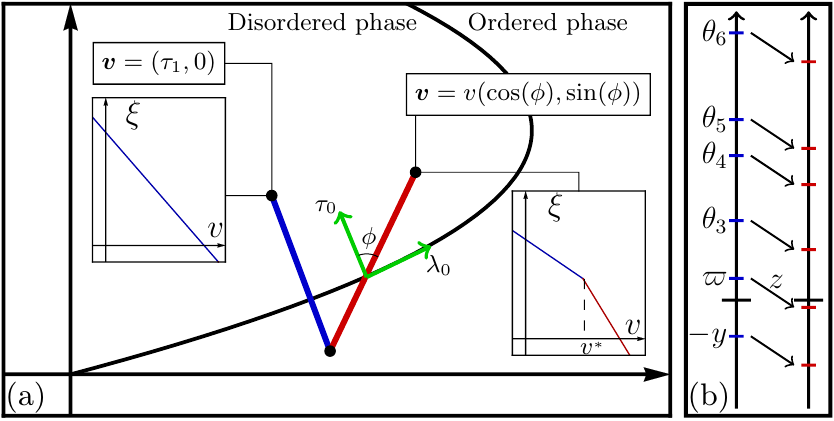}
\end{center}
\caption{(a): A generic phase diagram with two independent parameters (locally $\tau_0$ and $\lambda_0$) and a continuous phase transition. In the KZ scenario, the critical line is crossed transversally (blue line, left). As the phase transition is approached, the correlation length increases as $\xi \sim |\tau_0|^{-\nu}$ until adiabaticity breaks. At this point, the usual KZ scaling with the drive amplitude $v$, emerges $\xi \sim v^{-\nu/(1+z\nu)}$ (left inset, log-log). $z$ is the dynamical critical exponent. For more general drives involving a longitudinal component (red line, right), the correlation length may exhibit a different scaling $\xi \sim v^{-1/(z-\varpi)}$, which involves an irrelevant exponent $\varpi$. This behavior is realized when the drive amplitude is bigger than a crossover amplitude $v^*$ (right inset), that can be made to interpolate between $\infty$ and $0$ by changing the angle $\phi$, at which the phase boundary is crossed: thus, an irrelevant exponent is robustly observable in drive protocols with a strong enough longitudinal component. (b): Spectrum of equilibrium (blue, left) and nonequilibrium (right, red) critical exponents resulting from a linear drive. The drive shifts the whole spectrum down by $z$ and makes new relevant exponents out of irrelevant ones. This mechanism underlies the scaling of the correlation length with the drive amplitude described in (a) (see text).}
\label{fig_quenches}
\end{figure}

In this work, we demonstrate that irrelevant universal exponents can be turned into relevant ones by slowly driving the system in the vicinity of a second-order phase transition, classical or quantum, thus circumventing the need of fine tuning. This allows for the detection of these exponents in a robust way, and we provide a simple longitudinal drive protocol \cite{Divakaran2008} to do so in dynamical experiments (see \Fig{fig_quenches}{a}).

We describe the underlying physics qualitatively in \Sect{sec_basic_physical_mechanism} and derive our main result [detailed below \Eq{eq_stability_matrix}] in \Sect{sec_adiabatic_rg_flow}. In \Sect{sec_example} we choose a rather general model to illustrate our mechanism. In \Sect{sec_adiabaticity_breaking} we discuss how the breaking of adiabaticity is reflected in the \RG flow and how it can be seen to produce two different scales. We show how our mechanism can be observed in practice in \Sect{sec_observability}, and discuss the more general case of a polynomial drive in \Sect{sec_spectrum}. Finally, we illustrate the full mechanism on an exactly solvable model in \Sect{sec_solvable_model}.

\section{Basic physical mechanism}
\label{sec_basic_physical_mechanism}

Our result is obtained by a systematic reformulation and generalization of the \KZ scenario of diabatic decoupling in a time-dependent adiabatic \RG language \cite{tauber2014critical}. The \KZ mechanism \cite{kibble1976,Zurek1985,Damski2005,Zurek2005,Polkovnikov2006,Dziarmaga2010,Polkovnikov2011a,delCampo2014,Chandran2012,Huang2014,Nikoghosyan2016,Silvi2016} describes the behavior of a system when slowly ramping the parameters through a second-order phase transition: At a point in parameter space close enough to the phase transition, where the driving rate becomes comparable to the system's gap, the system crosses over from adiabatic to diabatic. Beyond that point, defects lead to a cutoff for the scaling of the correlation length (and far from equilibrium dynamics \cite{Biroli2010}). A quantitative prediction of this mechanism is the scaling of the correlation length at the crossover itself; according to the \KZ hypothesis, this scaling involves exclusively the exponents of the underlying equilibrium critical point.

Crucially, the quantitative predictions of the \KZ mechanism rely on identifying the point where adiabaticity breaks. The key idea of our RG approach is to first formulate adiabatic flow equations, where time enters only as a parameter, and then translate the breakdown of adiabaticity into an \RG language to connect to \KZ. As a first result of this approach, we demonstrate that the structure of the adiabatic flow equations reduces the \KZ hypothesis to the conventional scaling hypothesis: no independent information is encoded in them \cite{Zhong2006,Polkovnikov2011a,Baoquan2016}. This is because a slow drive only affects the large-scale physics (conversely, a fast drive acts at small scales and can produce new critical exponents \cite{DeSarkar2014,Monthus2017,Berdanier2018b,mathey2018a}). The most important new implication of our analysis is, however, based on the downshift of the spectrum of critical exponents (see \Fig{fig_quenches}{b}): Exponents that are irrelevant (positive sign) in the static problem now become relevant (negative sign). This implies that they must manifest themselves in large-scale physical observables. And indeed, we show that experiments probing \eg the correlation length can reveal irrelevant scaling exponents, provided that the slow ramp in parameter space crosses the phase boundary at a shallow enough angle (see \Fig{fig_quenches}{a}).

Our result is best understood by imagining the extreme case where the system is driven along the phase boundary. Then, the correlation length is always finite, but can get large if the system is held close enough to the phase boundary. In that case, the drive needs to be extremely slow for the dynamics to be adiabatic. With such a drive, we can imagine an alternative \KZ scenario where the system is driven multiple times along the phase boundary with the drive amplitude held fixed and the distance to the phase boundary decreasing. For large distance to the phase boundary the relaxation time is short and the system is adiabatic, and the latter increases as the former decreases. There is therefore a point where adiabaticity breaks since the drive amplitude is held fixed. The smaller the drive amplitude, the closer this point is to the phase boundary. As in the usual \KZ scenario, the correlation length scales with the drive amplitude at this point. The crucial difference, however, is that this longitudinal drive does not involve changing the distance to the phase boundary which is held fixed. It is instead an irrelevant parameter with a different scaling exponent that is driven. Although this parameter does not produce a diverging scale at equilibrium, it can still break adiabaticity when it is driven fast enough. It turns out that driving an irrelevant parameter in this way can actually produce a diverging length scale with its own scaling exponent. The \KZ scaling is thus modified as we show in detail below.

Since our scenario comprises both quantum and classical second-order phase transitions, it comes in timely for larger-scale quantum simulators such as the recently emerging Rydberg platforms that have already demonstrated the traditional \KZ scaling \cite{Keesling2018}. But, it is equally suitable for condensed matter \cite{Chae2012,Griffin2012,Grams2019}, ultracold atom \cite{Corman2014,Braun2015,Navon2015,Beugnon2017,Lienhard2018,Keesling2018,SafaviNaini2018,Liu2019}, superconducting \cite{Maniv2003,Monaco2009}, optical cavity \cite{Ducci1999,Baumann2011,Klinder2015,Kroeze2018}, or even hydrodynamic \cite{Casado2006} setups that offer the opportunity of slow parameter variations while exploring critical points \cite{delCampo2014}.

\section{Adiabatic \texorpdfstring{\RG}{RG} flow equations and exponent shift}
\label{sec_adiabatic_rg_flow}

In an adiabatic system, equilibration time scales are much shorter than those induced externally. Time then only enters through the parameters characterizing the system’s partition function. The full set of nonequilibrium adiabatic flow equations --- encoding the exact scaling dimensions of all operators --- is then obtained from its equilibrium counterpart by promoting the static parameters to time-dependent ones. They are
\begin{align}
 k\partial_k \bvec{\hat{g}} = \left[D_1+\eta \, D_2 \right] \bvec{\hat{g}} - z \, \bvec{\hat{g}}'\hat{t}+\bvec{\beta}(\bvec{\hat{g}}) \, .
 \label{eq_dimensionless_flow_equations_adiabatic}
\end{align}
This equation depends on time through the parameters, which are bundled in the vector $\bvec{\hat{g}}=\bvec{\hat{g}}(\hat{t}\,) = (\hat{g}^1(\hat{t}\,),\hat{g}^2(\hat{t}\,), \dots)$. We work in rescaled units where the canonical and anomalous components of the scaling dimensions are accounted for by the diagonal matrices $D_i$ and the anomalous dimension $\eta$. $k$ is a momentum \RG scale, and the prime denotes a time derivative. Here, time must be rescaled like any other parameter. This produces the second term on the right-hand-side because $\bvec{\hat{g}}$ depends on the rescaled time $\hat{t} = k^{z} t$, where $z$ is the dynamical critical exponent.

In principle, the above equation describes the renormalization of the full drive protocol through the entire time dependence of $\bvec{\hat{g}}$. We will, however, focus on linear drive protocols and simplify the problem accordingly. But first, we briefly sketch the derivation of the rescaled \RG flow equation \eq{eq_dimensionless_flow_equations_adiabatic}, starting from dimensionful adiabatic flow equations. In the adiabatic approximation, the dimensionful \RG flow equations are given by
\begin{align}
 & k\partial_k \bvec{g}(t) = \bvec{B}(\bvec{g}(t),Z(t),Q(t)) \, ,\nonumber \\
 & k\partial_k Z(t) = \eta(\bvec{g}(t),Z(t),Q(t)) \, Z(t) \, , \nonumber \\
 & k\partial_k Q(t) = z(\bvec{g}(t),Z(t),Q(t)) \, Q(t) \, .
 \label{eq_adiabatic_flow_equations}
\end{align}
$Z(t)$ and $Q(t)$ are the field strength and time rescaling factors respectively. At an \RG fixed point they scale as $Z\sim k^\eta$ and $Q \sim k^z$ and provide the anomalous dimension and the dynamical scaling exponent respectively. These equations are obtained from the dimensionful flow equations of the equilibrium system (which are computed in a standard way \cite{Zinnjustin2002a,tauber2014critical}) by promoting all the parameters to time-dependent ones. This is the definition of an adiabatic time dependence. \Eq{eq_adiabatic_flow_equations} is converted to its dimensionless form [\Eq{eq_dimensionless_flow_equations_adiabatic}] by rescaling the time and the coordinates of $\bvec{g}$ with appropriate powers of the \RG scale $k$, and the rescaling factors $Z$ and $Q$,
\begin{align}
 \hat{t} = Q t \, , && \bvec{\hat{g}}(\hat{t}\,) = k^{D_1} Z^{D_2} \bvec{g}(\hat{t}/Q)\, .
 \label{eq_rescaling}
\end{align}
We define
\begin{align}
\bvec{\beta}(\bvec{\hat{g}}) = k^{D_1} Z^{D_2} \bvec{B}(\bvec{\hat{g}}k^{-D_1}Z^{-D_2},Z,Q) \, ,
\end{align}
where the rescaling (choice of $D_1$ and $D_2$) is made in such a way that $k$, $Z$ and $Q$ drop out on the left-hand-side. Then, we obtain \Eq{eq_dimensionless_flow_equations_adiabatic} with the anomalous dimension and the dynamical critical exponents being functions of the rescaled parameters
\begin{align}
 \eta(\bvec{\hat{g}}) = \frac{k\partial_k Z}{Z} \, , && z(\bvec{\hat{g}}) = \frac{k\partial_k Q}{Q} \, .
\end{align}
Again, this is a standard procedure in RG theory \cite{Zinnjustin2002a,tauber2014critical}. Notably however, due to time being involved as an additional parameter to be rescaled in the presence of the slow drive, an additional term in \Eq{eq_dimensionless_flow_equations_adiabatic} (the middle term on the right-hand side) arises because the rescaled couplings now also depend on the cutoff through $Q$ [see \Eq{eq_rescaling}].

Now, to fix ideas, we choose the adiabatic time dependence of the parameters to be $\bvec{\hat{g}}(\hat{t}\,) = \bvec{\hat{g}}_0 + \bvec{\hat{g}}_1 \hat{t}$. Expanding \Eq{eq_dimensionless_flow_equations_adiabatic} to first order in $\hat{t}$ then leads to
\begin{align}
   & k \partial_k \bvec{\hat{g}}_0 = \left(D_1 + D_2 \eta\right) \bvec{\hat{g}}_0 + \bvec{\hat{\beta}}\, ,  \label{eq_dimensionless_flow_equations} \\
 & k \partial_k \bvec{\hat{g}}_1 = (D_1 + D_2 \eta - z ) \bvec{\hat{g}}_1 + \left[\frac{\partial \eta}{\partial \bvec{\hat{g}}} \hspace{-2pt} \cdot \hspace{-2pt} \bvec{\hat{g}}_1\right] D_2 \bvec{\hat{g}}_0+\frac{\partial \bvec{\hat{\beta}}}{\partial \bvec{\hat{g}}} \cdot \bvec{\hat{g}}_1 \, . \nonumber
\end{align}
This equation describes a stationary system (with parameters given by $\bvec{\hat{g}}_0$) in the presence of a small drive amplitude $\bvec{\hat{g}}_1$. The explicit time dependence of \Eq{eq_dimensionless_flow_equations_adiabatic} is traded for a doubling of the number of flowing parameters. The entire system's beta functions are obtained by joining \Eqs{eq_dimensionless_flow_equations} into a single vector, ${k\partial_k \vec{g} = (k\partial_k \bvec{\hat{g}}_0,k\partial_k \bvec{\hat{g}}_1) = \vec{\beta}(\vec{g})}$, with ${\vec{g} = (\bvec{\hat{g}}_0,\bvec{\hat{g}}_1)}$. The critical physics is then characterized by the equilibrium fixed point ${\vec{g}_{\text{fp}} = (\bvec{g}_{\text{fp}},\bvec{0})}$, that satisfies $\vec{\beta}(\vec{g}_{\text{fp}}) = 0$ and the flow close to it. In particular, if we define
\begin{align}
 \vec{G}(k)=\vec{g}(k)-\vec{g}_{\text{fp}} = (\bvec{\hat{G}}_0(k),\bvec{\hat{g}}_1(k)) \, ,
\end{align}
the components of $\vec{G}(k)$ are small in the critical region, and the flow is approximated as
\begin{align}
 k\partial_k \vec{G}(k) = M \vec{G}(k) \, ,
 \label{eq_linear_flow}
\end{align}
with $M$ is the Jacobian matrix of $\vec{\beta}(\vec{g})$, evaluated at the fixed point (see \App{app_structure_of_M}). We see from \Eq{eq_linear_flow} that the eigensystem of $M$ plays an important role at criticality.
Its eigenvalues are the critical exponents and its eigenvectors are used to define the corresponding parameters (see \App{app_scale_invariance}).

As a result of the adiabatic structure of the problem, the stability matrix $M$ has a relatively simple structure
\begin{align}
 M = \left( \begin{array}{cc} M_0 & X \\ 0 & M_1 \end{array}\right)\, .
 \label{eq_stability_matrix}
\end{align}
\hspace{-8pt} \begin{inparaenum}[(i)]
\item Its elements are square matrices of the same dimensionality as $\bvec{\hat{g}}$.
\item Its lower-left block vanishes identically because nonzero entries would signal that drive parameters ($\bvec{g}_1$, left-hand side) can be generated from equilibrium ones ($\bvec{g}_0$, right-hand side) alone.
\item Its diagonal blocks $M_0$ and $M_1$, represent the equilibrium and nonequilibrium critical physics, respectively. They are given in \App{app_structure_of_M}.
\item As a result of the adiabatic setup, the diagonal blocks are \emph{directly related}: $M_1 = M_0 -z$ [cf.~\Eq{eq_dimensionless_flow_equations}].
\item The upper-right block of $M$ describes the part of the renormalization of the equilibrium parameters that is generated by the drive. Although we find $X=0$ from \Eq{eq_dimensionless_flow_equations}, taking into account nonadiabatic correction leads to $X\neq0$ (see \App{app_structure_of_M}).
\end{inparaenum}

The critical exponents (including all subleading corrections) can be identified with the eigenvalues of $M$ [see \Eq{eq_linear_flow}]. The upper-triangular-block structure of \Eq{eq_stability_matrix}, implies that these are the eigenvalues of $M_0$ and $M_1$, and are thus independent of $X$. This leads to two key observations: First, the upper-left block is the only element of $M$ capable of producing independent exponents. This provides an \RG justification of the \KZ hypothesis: a slow drive does not produce any new critical exponents (see also \cite{Zhong2006,Baoquan2016}). Second, the relation $M_1 = M_0-z$ implies that the spectrum of critical exponents is doubled with the nonequilibrium copy shifted downwards by a factor $z$ (see \Fig{fig_quenches}{b}). Intriguingly, a parameter that is irrelevant at equilibrium, and characterized by a positive exponent $\varpi$, can thus be made relevant if $\varpi < z$. This means that this exponent must be associated with a \emph{diverging} length scale, \ie it must be observable macroscopically. This will be elaborated on below.

\section{Example}
\label{sec_example}

We now consider a concrete model. We show how the newly relevant operators can be observed in \Sect{sec_observability}. The general picture depends, however, on the precise form of the exponent spectrum and must be evaluated separately for each universality class. In particular, different combinations $z$ and drive protocol will lead to different scaling behaviors. We choose interacting $O(N)$ models. These provide common examples of well-known universality classes such as the Ising ($N=1$) \cite{Keesling2018} and Bose condensation ($N=2$) transitions. The spectrum of critical exponents of $O(N)$ models is estimated in \cite{juettner2017} for all values of $N$ and $d=3$. For example, the first four critical exponents of the Ising phase transition are found to be $\theta_1 \cong -1.54$, $\theta_2 \cong 0.66$, $\theta_3 \cong 3.18$, and $\theta_4\cong 5.91$. Moreover, $\theta_{m} = 2m-3 + \mathcal{O}(1/N)$ at large $N$ (with $m \geq 1$, integer). In all cases, $\theta_1=-y$ is the only relevant exponent and $\theta_{m>1}>0$ is irrelevant. The dynamical critical exponent $z$, can take different values that depend on the dynamical conservation laws \cite{tauber2014critical}. We choose the example of a dissipative order parameter (model A) where $z \cong 2 + 0.36 (N+2)/(N+8)^2$ for $d=3$ \cite{Zinnjustin2002a,tauber2014critical}. Then, a linear drive changes the relevance of a single parameter, and the four most relevant exponents,
\begin{align}
 \theta_1 = -y \, , && \theta_2 = \varpi \, , && -z-y \, , && -z+ \varpi \, ,
 \label{eq_critical_exp}
\end{align}
are associated with three relevant and one irrelevant parameters. The corresponding equilibrium system is controlled by the two parameters associated with the relevant and irrelevant exponents $-y$ and $\varpi$, which we refer to as $\tau_0$ and $\lambda_0$, respectively. We denote the corresponding drive parameters as $\tau_1$ and $\lambda_1$, which are associated to $-z-y$ and $\varpi-z$, respectively. See \App{app_scale_invariance} and \Eq{eq_projection} for a precise relation between these parameters and $\vec{g}$.

\section{Adiabaticity breaking and \texorpdfstring{\RG}{RG} analysis}
\label{sec_adiabaticity_breaking}

The \RG approach provides a direct connection between the physical breaking of adiabaticity and the emergence of a scale resulting from the drive: adiabaticity can be quantified by taking the ratio of the relaxation time\footnote{We choose the system's gap to be the first coordinate of $\bvec{g}_0$.} ${T_r=1/g^1_0}$, and the drive time-scale ${T^{\alpha} = |g^{\alpha}_0/g^{\alpha}_1|}$ for each parameter. Under \RG transformations, these quantities acquire a scale dependence
\begin{align}
 \epsilon^{\alpha}(k) = \frac{T_r(k)}{T^\alpha(k)} \cong \left|\frac{\hat{g}^{\alpha}_1(k)}{g_{\text{fp}}^{1} \, g_{\text{fp}}^{\alpha}}\right| \, .
  \label{eq_epsilon}
\end{align}
Adiabaticity is then signaled by a small such dimensionless ratio, and it gets broken once $|\bvec{\epsilon}(k)| \cong 1$.

To derive \Eq{eq_epsilon}, we focus on linear drive protocols.\footnote{Higher-order drives can, however, be taken into account by noting that they produce additional means of adiabaticity breaking. $\bvec{\epsilon}$ then acquires an additional index denoting the order of the drive $\epsilon^\alpha \rightarrow \epsilon^\alpha_i \sim \left|g^{\alpha}_i(k_0)\right|$ (with $i\geq 1$), and the following reasoning can be directly applied.} We write the dimensionful parameters as $\bvec{g} = \bvec{g}_0 + \bvec{g}_1 t$ [see \Eq{eq_rescaling}]. In terms of the microscopic parameters $\epsilon^{\alpha}$ is given by ${\epsilon^{\alpha} = \left|{g^{\alpha}_1}/{(g^1_0 g^{\alpha}_0)}\right|}$. The two time scales involved ($T_r$ and $T^{\alpha}$) are, however, strongly renormalized in the critical regime. Fluctuations are included by using cutoff-dependent parameters and choosing $k$ as small as possible:
\begin{align}
 \epsilon^{\alpha} = \left|\frac{g^{\alpha}_1(k)}{g^1_0(k) g^{\alpha}_0(k)}\right| =  \left|\frac{\hat{g}^{\alpha}_1(k)}{\left[g_{\text{fp}}^{1} +\hat{G}^1_0(k) \right] \left[g_{\text{fp}}^{{\alpha}} + \hat{G}^{\alpha}_0(k) \right] }\right| \, .
 \label{eq_epsilon_general}
\end{align}
We have inserted the rescaled variables in the second equality, and $\hat{G}_0^{\alpha}$ is the ${\alpha}^{\text{th}}$ component of $\bvec{\hat{G}}_0$. We consider the \RG flow close to an equilibrium fixed point where $\vec{g}_{\text{fp}} = (\bvec{g}_{\text{fp}},\bvec{0})$. For this reason, the coordinates of the distance from the fixed point can be identified with $\bvec{\hat{g}}_1$ in the nonequilibrium sector. Close to the fixed point we can expand the above equation to leading order in $\overrightarrow{G}$ and obtain \Eq{eq_epsilon}.

We emphasize that $\bvec{\epsilon}$ does not depend on the equilibrium couplings $\tau_0$ and $\lambda_0$, at criticality. Adiabaticity can only be broken if the system is driven. This is a consequence of the block-diagonal structure of $M$ (see \App{app_eigenvectors}). Indeed, expanding the solution of the linearized \RG flow onto the eigenvectors of $M$ eventually provides (see \App{app_transverse_longitudinal})
\begin{align}
 \bvec{\hat{g}}_1(k) = \tau_1 \left(\frac{\Lambda}{k}\right)^{y+z} \hspace{-0.1cm} \bvec{v}^1 + \lambda_1 \left(\frac{\Lambda}{k}\right)^{z-\varpi} \hspace{-0.1cm} \bvec{v}^2 \, ,
 \label{eq_rg_flow_g1}
\end{align}
with $\tau_1$ and $\lambda_1$ characterizing the two components of the drive and $\bvec{v}^{\alpha}$ the two eigenvectors of the equilibrium stability matrix [see \Eq{eq_longitudinal_transverse}]. $\tau_0$ and $\lambda_0$ do not enter in the flow of $\bvec{\hat{g}}_1$, which vanishes identically if the system is not driven.

From the point of view of the \RG, $|\bvec{\hat{g}}_1(k)|$ and therefore $|\bvec{\epsilon}(k)|$ increases as $k$ is lowered [see \Eq{eq_rg_flow_g1}]. To see how this is related to the emergence of a scale, we first recapitulate how the correlation length is extracted from the \RG under static equilibrium conditions. The \RG flow is initialized at a large momentum scale $k=\Lambda$ with the physical microscopic parameters, and evolves to effective macroscopic parameters as the \RG scale $k$ is lowered. In particular, for a near critical system, the distance from the fixed point is very small and $|{\tau}_0| \ll 1$. As a relevant parameter,\footnote{We use $(\hat{\tau}_0(k),\hat{\tau}_1(k),\hat{\lambda}_1(k))$ and $(\tau_0,\tau_1,\lambda_1)$ to denote the flowing couplings and the microscopic initial conditions respectively. In particular, we have $(\hat{\tau}_0(\Lambda),\hat{\tau}_1(\Lambda),\hat{\lambda}_1(\Lambda)) = (\tau_0,\tau_1,\lambda_1)$.} $\hat{\tau}_0(k)$ grows large under \RG transformations, and a scale $k_0$, emerges when $|\hat{\tau}_0 (k_0)| \approx 1$: ${k_0 \sim \xi^{-1}}$ provides the correlation length.

\begin{figure}[t]
 \begin{center}
\includegraphics[width=\columnwidth]{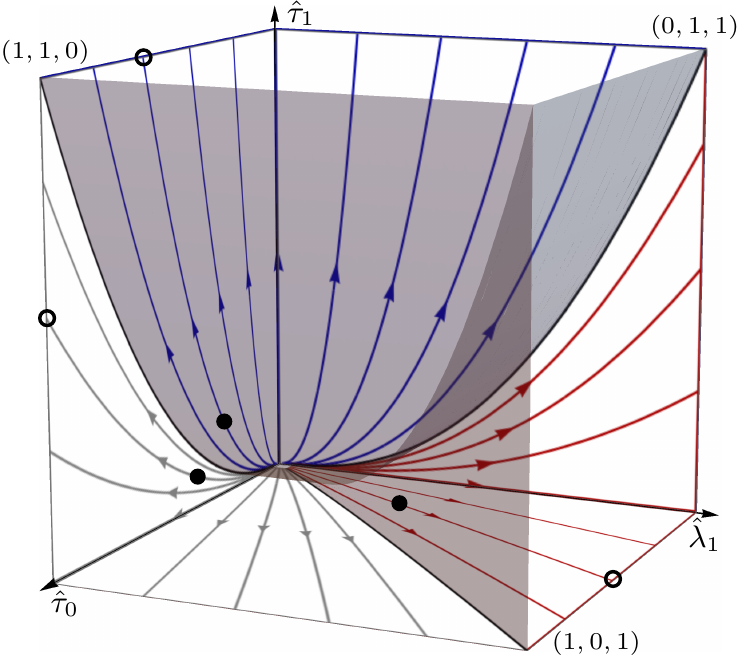}
 \end{center}
 \caption{RG flow close to the fixed point. In the presence of a drive, there are three relevant directions, and the RG flow can escape (empty circles) the fixed point from its microscopic initial conditions (filled circles) along different paths. There are three distinct scenarios, separated by the grey surface and identified by the biggest parameter at the point where the flow exits the cube: If the drive is slow enough (grey lines at the front), then the system is adiabatic and ${\xi \sim |\tau_0|^{-\nu}}$. When $|\tau_1|$ dominates (blue lines on the top), then ${\xi\sim\left|\tau_1\right|^{-\nu/(1+z\nu)}}$ (KZ scaling) and when $|\lambda_1|$ is big enough (red lines on the right) then ${\xi \sim |\lambda_1|^{-1/(z-\varpi)}}$ (new scaling).}
\label{fig_rg_flow}
\end{figure}

As demonstrated above, however, the system of adiabatic flow equations contains additional relevant directions emanating from the fixed point and associated to the slow drive. The three relevant directions for the case of $O(N)$ models are illustrated in \Fig{fig_rg_flow}{}. Along any of these directions, the flow will leave the critical scaling regime once the associated dimensionless parameters grow to $\mathcal O (1)$. We discuss the physics of the flow initialized close to the fixed point along the primitive axes of the coordinate system first:
\begin{itemize}
 \item The flow along the $\hat{\tau}_0$ axis corresponds to the case of an undriven system, $\tau_1 = \lambda_1 =0$. In that case, the entire flow is controlled by the equilibrium critical exponent and we have $|\vec{g}(k)-\vec{g}_{\text{fp}}| \sim \hat{\tau}_0(k) \sim \tau_0 \, k^{-y}$, with ${\tau_0 = (T-T_c)/T_c}$ the reduced temperature. The correlation length --- extracted from the solution of the flow equation where $|\hat{\tau}_0(k_0)| \approx 1$ --- scales as $\xi \sim \tau_0^{-\nu}$ with $\nu = 1/y$. In particular we recover the Ising exponent $\nu \cong 0.65$, when $y \cong 1.54$. This reproduces the usual link between the critical exponents \Eq{eq_critical_exp}, and the scaling of the correlation length.
 \item The nonequilibrium drive provides two additional scaling regimes, defined by the plane ${\tau}_0 =0$. Crucially, the emergence of a scale along these directions [\iec $\hat{\tau}_1(k)$ or $\hat{\lambda}_1(k)$ becoming $\mathcal{O}(1)$] coincides with the breaking of adiabaticity, as can be read off from \Eq{eq_epsilon}. Indeed, the coordinates of $\bvec{\hat{g}}_1(k)$ are linear combinations of $\hat{\tau}_1(k)$ and $\hat{\lambda}_1(k)$ (see \Eq{eq_rg_flow_g1} and \App{app_transverse_longitudinal}) and the fixed point parameters $\bvec{g}_{\text{fp}}$ are $\mathcal{O}(1)$. The observable scaling is eventually determined by the relative amplitude of the different components of $\bvec{g}_1$: When the flow follows the $\hat{\tau}_1$ axis ($\tau_0 = \lambda_1 =0$), we have $|\vec{g}(k)-\vec{g}_{\text{fp}}| \sim |\tau_1| \, k^{-y-z}$. Then, we can extract $k_0$ just as before, and recover the usual \KZ scaling $\xi \sim |\tau_1|^{-\nu/(1+z\nu)}$. Conversely, when the flow follows the $\hat{\lambda}_1$ axis, we get $|\vec{g}(k)-\vec{g}_{\text{fp}}| \sim |\lambda_1| \, k^{\varpi-z}$, and $\xi \sim |\lambda_1|^{-1/(z-\varpi)}$. The new scaling exhibiting the irrelevant exponent $\varpi$ takes place.
\end{itemize}

Thus, all together there are three possible ways for a scale to emerge and two ways to break adiabaticity. We represent the crossovers from the different scaling regimes as gray surfaces in \Fig{fig_rg_flow}{} with the scaling in each region determined by its encompassed axis. The system is adiabatic in the lower-left corner only, where we have $\xi \sim |\tau_0|^{-\nu}$. Conversely, if the drive is strong (or if $|\tau_0|$ is small) enough, the system is diabatic. Then, the correlation length at the crossover from adiabatic to diabatic behaves according to one of the two nonequilibrium relevant scalings. Crucially, there are two possible exponents: ${\nu_1 = \nu/(1+z\nu)}$ is the usual \KZ exponent, and ${\nu_2 = 1/(z-\varpi)}$ is a new one containing the irrelevant exponent $\varpi$. We emphasize that the difference in exponents is significant: for example, for model A at $N^{-1}=0$, $\nu_1 = 1/3$ and $\nu_2 = 1$.

We close this section by using \Eq{eq_epsilon_general} to extract the scaling of the crossover between the adiabatic and diabatic cases. We have seen that $|\bvec{\epsilon}|$ increases as $k$ decreases and that adiabaticity is broken if $|\bvec{\epsilon}(k_0)| \sim 1$. There are two possibilities: If $|\tau_0|$ is large enough for $|\bvec{\hat{G}}_0(k)|$ to become of order $1$ before $|\bvec{\hat{g}}_1(k)|$, then the denominator of \Eq{eq_epsilon_general} becomes order $1$ before its numerator has a chance to become large and $|\bvec{\epsilon}(k_0)|\ll 1$. The system is adiabatic and the correlation length scales as $\xi \sim |\tau_0|^{-\nu}$. When $|\tau_0|$ is smaller, it is $|\bvec{\hat{g}}_1(k)|$ that is of order $1$ at $k_0$. Then, adiabaticity is broken and the correlation length scales as ${\xi \sim |\bvec{g}_1|^{-\nu_{1,2}}}$. In other words, adiabaticity is broken when the scaling of $\xi$ with $\tau_0$ saturates. Equating the two scales $|\tau_1|^{-\nu} = |\bvec{g}_1|^{-\nu_{1,2}}$, provides different crossovers for the two drive protocols: the usual \KZ scaling emerges ($|\tau_1|\gg |\lambda_1|$) when ${|\tau_1| \gtrsim |\tau_0|^{1+z\nu}}$ and the new scaling is visible when ${|\lambda_1| \gtrsim |\tau_0|^{\nu(z-\varpi)}}$. Additionally, we find that $\bvec{\epsilon}$ scales as ${|\bvec{\epsilon}| \sim |\tau_0|^{-\nu/\nu_i}}$ in the adiabatic regime and thus diverges as $\tau_0 \rightarrow 0$. Indeed, in that regime we have ${k_0 \sim |\tau_0|^{\nu}}$. Inserting this in \Eq{eq_rg_flow_g1} provides ${|\bvec{\hat{g}}_1(k_0)| \sim |\tau_0|^{-\nu/\nu_i}}$ with $\nu_i$ chosen according to the nature of the drive protocol.

\section{Observability and robustness}
\label{sec_observability}

We now connect these \RG findings to concrete observables. To this end, we start from the drive protocol illustrated in the phase diagram of \Fig{fig_quenches}{a} (in red, on the right) and parametrize
\begin{align}
 \bvec{\hat{g}}_1= \left(\begin{array}{c} \tau_1 \\ \lambda_1 \end{array}\right) = v \left(\begin{array}{c} \cos(\phi) \\ \sin(\phi) \end{array}\right) \, .
 \label{eq_def_v}
\end{align}
$v$ denotes the drive amplitude and $\phi$ its direction. $\tau_1$ and $\lambda_1$ are respectively associated with driving the system across and along the phase boundary because they are analogous to (\ie they scale with the shifted exponent of) $\tau_0$ and $\lambda_0$ (see \Fig{fig_quenches}{a}, green arrows and \App{app_eigenvectors}). This provides the following interpretation: When the drive is perpendicular to the phase boundary ($\phi=0$), it is characterized by $\tau_1$ and the scaling is ${\xi \sim v^{-\nu_1}}$. In that case we recover the usual \KZ scaling. When the system is driven along the phase boundary ($\phi=\pi/2$), $\lambda_1$ takes over, and the irrelevant critical exponent $\varpi$ is visible.

\begin{figure}[t]
 \begin{center}
\includegraphics[width=\columnwidth]{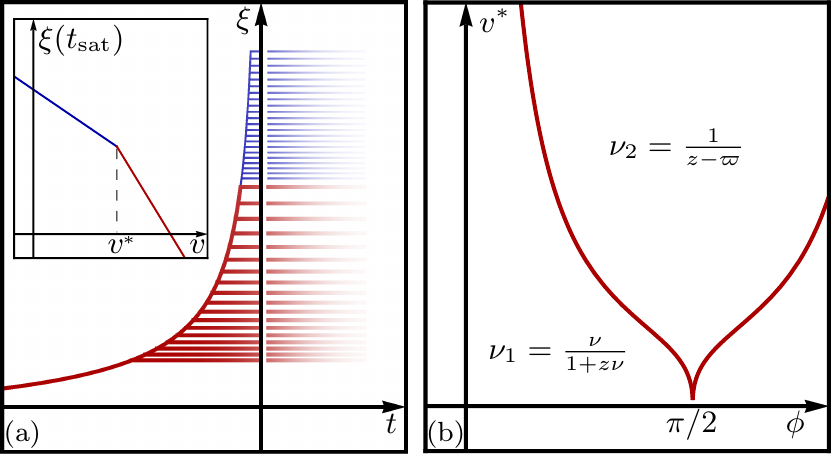}
 \end{center}
 \caption{(a): Time dependence of the correlation length for different drive amplitudes.  As the system is driven across the phase transition $\xi$ increases as $\xi \sim |\tau_0|^{-\nu}$ until adiabaticity breaks at $t=t_{\text{sat}}$. At this time, $\xi$ scales with the drive amplitude as $\xi(t_\text{sat}) \sim v^{-\nu_i}$. There are two scaling regimes (a, inset, log-log plot). For $v<v^*$ (transversal drive), the usual KZ scaling is visible $\nu_i = \nu/(1+z\nu)$, while for $v>v^*$ (longitudinal drive) the new exponent $\nu_i = 1/(z-\varpi)$ emerges. (b): The crossover amplitude $v^*$, depends on the drive direction (parametrized by the angle $\phi$), and can be made arbitrarily small by choosing $\phi$ close enough to $\pi/2$.}
\label{fig_robustness}
\end{figure}

Crucially, a fine tuning of $\phi$ is not necessary, and both scalings can be observed for an extended set of parameters. Indeed, we find that the scaling of $\xi$ with $v$ displays two regimes separated by a crossover drive amplitude $v^*$. The \KZ and the new exponents are visible for $v$ smaller and bigger than $v^*$ respectively. Notably, $\phi$ provides a means to bring either one of these regimes forward because $v^*$ interpolates from infinity to zero as $\phi$ is tuned from zero to $\pi/2$ (See \Fig{fig_robustness}{b}).

The dependence of the crossover drive amplitude on $\phi$ can be estimated for angles close to zero and to $\pi/2$ [see \Eq{eq_cross_over_scaling}]. This is readily understood by viewing the system as depending on $\tau_0$ with a fixed drive amplitude. For $\tau_0$ large enough (the system is adiabatic and) the correlation length scales as $\xi \sim \tau_0^{-\nu}$. We see that $\xi$ is bigger for smaller $\tau_0$. It can however not be arbitrarily large since adiabaticity eventually breaks if $\tau_0$ is decreased while $v$ is held fixed. This happens when the correlation length reaches the smallest of the two underlying scales
\begin{align}
 \xi_{1} \sim v^{-\nu/(1+z\nu)} \, , && \xi_{2} \sim v^{-1/(z-\varpi)} \, .
\end{align}
See \cite{Huang2014} where a similar picture emerges. For $v$ small enough, $\xi_1$ is always smaller than $\xi_2$ because $\nu/(1+z\nu)<1/(z-\varpi)$. This means that the usual \KZ scaling is always visible when $v\rightarrow 0$. A crossover emerges because $\xi_1$ is bigger than $\xi_2$ when $v$ is large enough. $v^*$ is obtained by equating the two scales
\begin{align}
 |v^* \cos(\phi)|^{1/(y+z)} = |v^* \sin(\phi)|^{1/(z-\varpi)} \, .
\end{align}
See \Eq{eq_def_v}. Its dependence on $\phi$ can be estimated for angles close to zero and $\pi/2$ (see \Fig{fig_robustness}{b})
\begin{align}
 & v^* \sim \phi^{-(z \nu +1)/(\varpi \nu +1)} && \text{for } && \phi \ll 1 \nonumber \\
 & v^* \sim \left|\phi-\pi/2\right|^{\nu(z-\varpi)/(1+ \nu \varpi)} && \text{for } && \phi \cong \pi/2 \, .
 \label{eq_cross_over_scaling}
\end{align}
We see that $v^*$ vanishes when $\phi = \pi/2$ and is small when $\phi$ is close enough to $\pi/2$. The longitudinal scaling emerges when the system is driven at a shallow angle with the phase boundary.

Finally, we connect our quasistatic findings to an experimental procedure where the system is driven across the phase transition, taking into account the instantaneous value of the distance from the phase transition $\tau(t) = \tau_0 + v \cos(\phi) t$ (see \Fig{fig_quenches}{a}). Adiabaticity inevitably breaks since $\tau(t)$ crosses zero, and the above results are applied at this moment. In particular, $v^*$ can be made arbitrarily small by crossing the phase boundary at a shallow enough angle.

The above scaling phenomenology is not bound to the correlation length showcased so far. It will emerge in all typical \KZ observables, such as the relaxation time $T_r \sim \xi^z \sim v^{-z\nu}$, or the defect density (if these are created as the system crosses the phase boundary), which is evaluated from the correlation volume $n_{\text{d}} \sim \xi^{-d}\sim v^{d\nu_i}$ (see \egc \cite{Chandran2012}). Both carry information on the irrelevant exponent for $v>v^*$.

\section{Spectrum of irrelevant exponents}
\label{sec_spectrum}

In principle, the whole spectrum of irrelevant exponents is accessible through an appropriate slow drive. For example, a higher-order drive $\bvec{\hat{g}} = \bvec{\hat{g}}_0 + \bvec{\hat{g}}_2 \hat{t}^{\,2}/2$ produces a doubly shifted copy of the spectrum of critical exponents: $M_2 = M_0 - 2z$ (see \App{app_structure_of_M}). In particular the newly relevant exponent $\theta_3-2z$ produces a scaling regime with $\xi \sim v^{-1/(2 z-\theta_3)}$ and enables the observation of the next irrelevant exponent $\theta_3$ by adjusting $\bvec{\hat{g}}_2$ longitudinally to the phase boundary. In that case, however, two exponents are made relevant since the first irrelevant exponent is also shifted by $2z$, and two irrelevant couplings must be taken into account. The second order drive must follow the phase boundary in a direction that favors the observation of $\theta_3$ over $\varpi$ (see \App{app_transverse_longitudinal}). As this procedure is iterated, the dimensionality of the required parameter space grows by one for each newly relevant critical exponent. The general principle is as follows: Consider an equilibrium phase diagram with $r+1$ axes and an $r$-dimensional critical surface, \iec with one direction crossing the transition (equilibrium relevant parameter, transversal direction) and $r$ irrelevant (longitudinal) directions. Then, a polynomial drive of order $r$ can activate the first $r$ irrelevant critical exponents.

We point out that in the case of a transversal polynomial drive of order $r$, the most relevant operator, which is $-y-rz$, immediately leads to the known scaling ${\xi \sim v^{-\nu/(1+z r \nu)}}$ \cite{Sen2008,DeGrandi2010a,DeGrandi2010b,Polkovnikov2011a}. Moreover, our approach can also be applied to the problem of adiabaticity restoration with a symmetry-breaking field (see, \egc \cite{Rams2019,Rysti2019}). These fields are relevant and produce a breaking of scale invariance with an additional negative critical exponent $-y_2$. In particular, driving the system through its critical point with such a field will produce a new scaling with $\xi \sim v^{-1/(z+y_2)}$.

\section{Solvable model}
\label{sec_solvable_model}

In this section we illustrate our result with a non-interacting toy model, which is not particularly realistic but exactly solvable. In particular, we recover our result without using the \RG. We consider the fluctuating steady state of the Langevin dynamics in $d$ spatial dimensions,
\begin{align}
 \partial_t \psi = - m \, \psi - \nabla^4 \psi + Q \, \nabla^6 \psi + \zeta \, .
 \label{eq_langevin_equation}
\end{align}
$\psi$ is a real space and time dependent field and $\zeta$ a Gaussian white noise with $\langle \zeta(t,x) \zeta(t',x')\rangle = 2 \delta(t-t') \delta(x-x')$. We scale the field $\psi$ such that the temperature that usually appears in the noise correlator is set to one. We start with the application of our \RG analysis. The renormalization of \Eq{eq_langevin_equation} is simple because there is no interaction. The flow of $m$ and $Q$ are given by
\begin{align}
 k\partial_k \hat{m} = -4 \, \hat{m} \, , && k \partial_k \hat{Q} = 2 \, \hat{Q} \, .
\end{align}
There is a fixed point at $\hat{m}=\hat{Q}=0$ with one relevant ($y=4$) and one irrelevant ($\varpi=2$) exponent. Furthermore, the dynamical critical exponent is given by $z=4$. In the presence of a linear drive
\begin{align}
 m(t) = m_0+ m_1 t \, , && Q(t) = Q_0 + Q_1 t \, ,
 \label{eq_drive}
\end{align}
our work predicts the equilibrium ($\nu = 1/4$), transversal and longitudinal scaling to be
\begin{align}
 \xi \sim |m_0|^{-1/4} \, , && \xi \sim v^{-1/8} \, , && \xi \sim v^{-1/2} \, ,
 \label{eq_rg_scaling}
\end{align}
respectively. The drive amplitude is given in terms of the dimensionless parameters at the beginning of the \RG flow: $v^2 = (m_1/\Lambda^8)^2+(Q_1/\Lambda^2)^2$.

We now reproduce these results from a less formal, independent argumentation. Although \Eq{eq_langevin_equation} can be solved exactly with the parameters of \Eq{eq_drive}, this is actually not necessary. The equal-time correlation function,
\begin{align}
 \langle\psi(t,r) \psi(t,0) \rangle = G(t,r;m_0,Q_0,m_1,Q_1) \, ,
 \label{eq_correlation_function}
\end{align}
depends on space $r=|(r_1,r_2,\dots,r_d)|$, time $t$, and the system parameters. We apply dimensional analysis and write the above equation in terms of dimensionless quantities. The different elements of \Eq{eq_correlation_function} have the following dimension 
\begin{align*}
& [r] = L\, , && [t] = L^4 \, , && [\psi] = L^{(4-d)/2} \, , && \nonumber \\
& [m_0] = L^{-4} \, , && [Q_0] = L^{2} \, , && [m_1] = L^{-8} \, ,&&  \hspace{-8pt} [Q_1] = L^{-2} \, .
\end{align*}
These dimensions can all be written as powers of a single length $L$, because the parameter in front of $\nabla^4$ in \Eq{eq_langevin_equation} and the temperature have both been set to $1$. This dimensional analysis implies that the equation of motion in terms of the rescaled parameters
\begin{align*}
& \hat{r} = \frac{r}{L}\, , && \hat{t} = \frac{t}{L^{4}} \, , && \hat{\psi} = \frac{\psi}{L^{(4-d)/2}} \, , && \nonumber \\
& \hat{m}_0 = m_0 L^{4} \, , && \hat{Q}_0 = \frac{Q_0}{L^{2}} \, , && \hat{m}_1 = m_1 L^{8} \, , && \hspace{-8pt}  \hat{Q}_1 = Q_1 L^{2} \, ,
\end{align*}
is identical to \Eq{eq_langevin_equation}, with $L$ not appearing explicitly. In particular this implies that
\begin{align}
 \langle\psi(t,r) \psi(t,0) \rangle = L^{4-d} \langle \hat{\psi}(\hat{t},\hat{r}) \hat{\psi}(\hat{t},0) \rangle \, ,
\end{align}
for any choice of $L$. We now insert the rescaled variables, choose $L=r$ and identify $G$ on both sides of the above equation
\begin{align}
 & G(t,r;m_0,Q_0,m_1,Q_1) \nonumber \\
 & \qquad = r^{4-d} \, G\left(\frac{t}{r^4},1;m_0 r^4,\frac{Q_0}{r^2},m_1 r^8,Q_1 r^2\right) \nonumber \\
 & \qquad = r^{4-d} \, \hat{G}\left(\frac{t}{T_r};\frac{r}{\xi_{m_0}},\frac{r}{\xi_{Q_0}},\frac{r}{\xi_{m_1}},\frac{r}{\xi_{Q_1}}\right)\, ,
\end{align}
with $T_r = r^4$, and
\begin{align*}
 & \xi_{m_0} = m_0^{-1/4} \, , && \xi_{Q_0} = Q_0^{1/2} \, , \nonumber \\
 & \xi_{m_1} = m_1^{-1/8} \, , && \xi_{Q_1} = Q_1^{-1/2} \, .
\end{align*}
$\hat{G}$ is a dimensionless function that does not depend on the couplings.

We see that the four couplings $m_0$, $m_1$, $Q_0$ and $Q_1$ together with the ramp time $t$, each produce a different scale. Furthermore, the scale produced by the ramp time $t/T_r = [t^{1/4}/r]^4$, is $\xi_t = t^{1/4}$. We choose small values of $|t|$ because we are interested in the near critical physics. Close to the fixed point (where all the couplings vanish and $|t|$ is small) we find that, in accordance with the \RG analysis,
\begin{itemize}
 \item $\xi_t$ is small.
 \item $Q_0$ produces a scale that asymptotically vanishes as we approach the fixed point. It does not diverge and therefore is not visible on large spatial scales. $Q_0$ is an irrelevant parameter.
 \item The scale of $m_0$ diverges with the critical exponent $\nu = 1/4$.
 \item The scale of $m_1$ diverges with the usual \KZ exponent $\nu_1=\nu/(1+z\nu)= 1/8$.
 \item The scale of $Q_1$ diverges with the predicted scale $\nu_2=1/(z-\varpi) = 1/2$.
\end{itemize}

We now set $Q_0 = t=0$ for simplicity and, as in the general analysis, we introduce $m_1 = \Lambda^8 v \cos(\phi)$, $Q_1 = \Lambda^2 v \sin(\phi)$. Then, the correlation function behaves as a power law ($G \sim r^{4-d}$) as long as $r \ll \text{Min}(\xi_{m_0},\xi_{m_1},\xi_{Q_1})$ and then decays exponentially. We can therefore identify the correlation length with the smallest of the three observable scales, $\xi = \text{Min}(\xi_{m_0},\xi_{m_1},\xi_{Q_1})$. For a fixed value of $\phi$ we extract $v^*$ as the value of $v$ where $\xi_{m_1} = \xi_{Q_1}$. It is given by $[v^* \cos(\phi)]^{1/8} = [v^* \sin(\phi)]^{1/2}$. Then, the three regimes that we discuss in our paper emerge naturally:
\begin{itemize}
 \item If $m_0$ is large enough, $\xi_{m_0}$ is the smallest scale and $\xi = m_0^{-1/4}$. The system is adiabatic and exhibits equilibrium scaling.
 \item If $m_0$ is small and $v<v^*$, then $\xi_{m_1}$ is the smallest scale and we see the usual \KZ scaling $\xi \sim v^{-1/8}$.
 \item If $m_0$ is small and $v>v^*$, then $\xi_{Q_1}$ takes over and we see the new scaling $\xi \sim v^{-1/2}$.
\end{itemize}

We conclude with a remark concerning our choice of model. We have chosen this somewhat unusual equation (with $\nabla^4$ and $\nabla^6$) because it illustrates our result very cleanly. Indeed, with a more typical model (with $\nabla^2$ and $\nabla^4$), our \RG analysis would still be applicable, but we would have to resort to a quadratic drive because $z=\varpi$. In this case, the linear drive makes the irrelevant parameter marginal (vanishing critical exponent), not relevant.

\section{Conclusion}

In an \RG language, the \KZ mechanism allows one to observe relevant critical exponents by driving along a relevant scaling direction, \iec transversally to the phase boundary. We find that irrelevant exponents can be made relevant, and therefore observable, by driving longitudinally to the phase boundary. The observability is robust, persisting to the presence of weak transversal drive components. The quantitative difference between the exponents is quite significant if, as usually the case, the full critical exponents are close enough to the canonical ones. It therefore stands to reason that the mechanism uncovered here may underlie some of the difficulties in determining critical exponents in \KZ experiments \cite{Monaco2009}, and may help to foster progress in this direction.

\section{Acknowledgments}

We thank A. Chiocchetta, L. Corman, B. Delamotte, C. Grams, J. Hemberger, M. Kastner, B. Ladewig, P. H. M. van Loosdrecht, G. Morigi, A. Rosch, U. Schneider, F. Sekiguchi and R. Versteeg for useful and inspiring discussions. We acknowledge support by the Institutional Strategy of the University of Cologne within the German Excellence Initiative (ZUK 81), by the funding from the European Research Council (ERC) under the Horizon 2020 research and innovation program, Grant Agreement No. 647434 (DOQS), and by the DFG Collaborative Research Center (CRC) 1238 Project No. 277146847 - project C04. This research was supported in part by the National Science Foundation under Grant No. NSF PHY-1748958.

\appendix

\section{Structure of the stability matrix}
\label{app_structure_of_M}

In this appendix we show how the structure of the stability matrix that is described in the main text emerges from the adiabatic approximation. In particular, we recover \Eq{eq_stability_matrix}, provide an explicit expression for $M_0$ and $M_1$, and identify the role of the adiabatic approximation for each element.

We now formally consider the full \RG flow equations without approximation, but use a representation of the flowing parameters that is particularly suited to the adiabatic approximation,
\begin{align}
 \bvec{\hat{g}}(\hat{t}\,) = \sum_i \bvec{\hat{g}}_i \frac{\hat{t}^i}{i!} \, .
 \label{eq_gt_expansion}
\end{align}
In general, the dimensionless \RG flow equations take the form
\begin{align}
 k\partial_k \bvec{\hat{g}}(\hat{t}\,) = \left(D_1+\eta[\bvec{\hat{g}}] D_2 \right) \bvec{\hat{g}} - z[\bvec{\hat{g}}] \bvec{\hat{g}}' \hat{t}+\bvec{\beta}(\bvec{\hat{g}}) + \bvec{L}(\bvec{\hat{g}})[\bvec{\hat{g}}'] \, .
 \label{eq_dimensionless_flow_equations_noapprox}
\end{align}
Every term in the above equation depends on the rescaled time $\hat{t}$ explicitly, or through $\bvec{\hat{g}}(\hat{t}\,)$ and its time-derivatives. The square brackets denote a functional dependence of the form $X[f] = X(f(t),f'(t),f''(t),\dots)$. The last term on the right-hand-side of \Eq{eq_dimensionless_flow_equations_noapprox} contains the diabatic loop corrections. It vanishes in the absence of drive $\bvec{L}(\bvec{\hat{g}})[\bvec{0}] = 0$. The remaining diabatic effects are included in the functional dependence of $\eta$ and $z$ on $\bvec{\hat{g}}$. In the adiabatic approximation we recover \Eq{eq_dimensionless_flow_equations_adiabatic} by setting $\bvec{L}(\bvec{\hat{g}})[\bvec{\hat{g}}'] = 0$, $\eta[\bvec{\hat{g}}] = \eta(\bvec{\hat{g}})$ and $z[\bvec{\hat{g}}] = z(\bvec{\hat{g}})$.

In the presence of an arbitrary drive \Eq{eq_gt_expansion}, the \RG flow equations of $\bvec{\hat{g}}_i$ are obtained by Taylor expanding \Eq{eq_dimensionless_flow_equations_noapprox} in powers of $\hat{t}$,
\begin{align}
 & k\partial_k \bvec{\hat{g}}_i = \left. \frac{\partial^i}{\partial\hat{t}^i} k\partial_k \bvec{\hat{g}}(\hat{t}\,) \right|_{\hat{t}=0} \, ,
\end{align}
with \Eq{eq_dimensionless_flow_equations_noapprox} inserted on the right-hand-side. Then, the different blocks of the stability matrix are obtained by differentiating the above equation with respect to $\bvec{\hat{g}}_i$ and evaluating at $\bvec{\hat{g}}(\hat{t}\,) = \bvec{g}_{\text{fp}}$,
\begin{align}
 M_{ij} = \left. \frac{\partial^{i+1}}{\partial \bvec{\hat{g}}_j \partial\hat{t}^i} k\partial_k \bvec{\hat{g}}(\hat{t}\,) \right|_{\hat{t}=0,\bvec{\hat{g}}_0 = \bvec{g}_{\text{fp}},\bvec{\hat{g}}_{n>0} = 0} \, ,
 \label{eq_stability_matrix_general}
\end{align}
with \Eqs{eq_dimensionless_flow_equations_noapprox} and \eq{eq_gt_expansion} inserted on the right-hand-side.

It is easier to evaluate \Eq{eq_stability_matrix_general} if we start with the derivatives with respect to $\bvec{\hat{g}}_i$. Indeed, evaluating the parameters at $\bvec{\hat{g}}(\hat{t}\,) = \bvec{g}_{\text{fp}}$ simplifies greatly the time dependence since $\bf{g}_{\text{fp}}$ does not depend on time. We find that the stability matrix is block upper triangular ($M_{ij}=0$ if $j<i$) with the different blocks being square matrices with the dimension of the space of equilibrium parameters. They are
\begin{align}
 & M_{ii} = D_1+\eta D_2 + D_2 \bvec{g}_{\text{fp}} \frac{\partial \eta}{\partial \bvec{\hat{g}}_{0}} - i z  + \frac{\partial \bvec{\beta}}{\partial \bvec{\hat{g}}} \nonumber \\
 & M_{i<j} = \frac{\partial \eta}{\partial \bvec{\hat{g}}_{j-i}} D_2 \bvec{g}_{\text{fp}} + \frac{\partial \bvec{L}}{\partial \bvec{\hat{g}}_{j-i}} \, .
 \label{eq_stability_matrix_blocks}
\end{align}
We use Latin indices ($i,j=0,1,\dots$) to denote the different drive sectors. We give an explicit expression for upper-left block $M_{00}$ that couples $\bvec{g}_0$ to itself (equilibrium physics):
\begin{align*}
 [M_{00}]_{\alpha \beta} = [D_1]_{\alpha \beta} + \eta [D_2]_{\alpha \beta} + [D_2 \bvec{g}_{\text{fp}}]_{\alpha} \hspace{-0.1cm} \left.\frac{\partial \eta}{\partial \hat{g}_{\beta}}\right|_{\bvec{g}_{\text{fp}}} \hspace{-0.3cm} + \hspace{-0.1cm} \left.\frac{\partial \hat{\beta}_{\alpha}}{\partial \hat{g}_{\beta}}\right|_{\bvec{g}_{\text{fp}}} \hspace{-0.2cm} .
\end{align*}
We use Greek indices [with $\alpha,\beta = 1,2,\dots,\text{dim}({\bvec{g}})$] to denote the coordinates within each block. The above equation makes the notation used in \Eq{eq_stability_matrix_blocks} clear. We see that all the diagonal blocks [referred to as $M_0$ and $M_1$ in \Eq{eq_stability_matrix}] are determined from the equilibrium block
\begin{align}
 M_{ii} = M_{00} - i \, z \, .
 \label{eq_shift_exponents}
\end{align}

We see from \Eq{eq_stability_matrix_blocks}, that the diagonal blocks of $M$ are not affected by the adiabatic approximation. The off-diagonal blocks, however, are because they contain on $L(\hat{g})[\hat{g}']$ and the derivatives of $\eta[\hat{g}]$ with respect to the time-derivatives of $\bvec{\hat{g}}$.

\section{Emergence of scales}
\label{app_scale_invariance}

In this appendix we discuss the emergence of scales obtained through the \RG. In particular, we show how these are related to the critical exponents and the eigenvectors of the stability matrix. This in turn provides precise definitions for the parameters $\tau_0$, $\lambda_0$, $\tau_1$, and $\lambda_1$ that were introduced in the main text. Moreover, we show that the upper-triangular structure of $M$ leads to a classification of the different drive protocols as longitudinal and transversal when the drive protocol is truncated to a finite order
\begin{align}
 \bvec{\hat{g}}(\hat{t}\,) = \sum_{i=0}^r \bvec{\hat{g}}_i \frac{\hat{t}^i}{i!} \, .
 \label{eq_gt_expansion_truncated}
\end{align}
$\bvec{\hat{g}}$ and $\bvec{\hat{g}}_i$ (with $i=0,1,\dots,r$) are vectors that can be represented on the system's phase diagram. They have as many coordinates as there are parameters in the equilibrium system.

\subsection{Eigenvectors of \texorpdfstring{$M$}{M} and diagonal parameters}
\label{app_eigenvectors}

We start by relating the eigensystem of the stability matrix to the microscopic parameters. The structure of the stability matrix enables its diagonalization in terms of the eigensystem of the equilibrium stability matrix $M_{0}$, which we denote as as
\begin{align}
 M_{0} \bvec{v}^{\alpha} = \theta_{\alpha} \bvec{v}^{\alpha} \, ,
 \label{eq_eigensystem}
\end{align}
with $\alpha\geq 1$ integer. $\bvec{v}^\alpha$ has the same dimension as $\bvec{g}$ and $\alpha = 1,2,\dots$ runs from $1$ to the dimension of $\bvec{g}$ as well. In the notation of the main text we have ${\theta_1 = -y = -1/\nu}$ and ${\theta_2 = \varpi}$.

The eigenvalues of the full stability matrix $M$, are given by the eigenvalues of its diagonal blocks because $M$ is upper-triangular. Moreover, the general relation between the different diagonal blocks \Eq{eq_shift_exponents}, implies that the spectrum of $M$ comes in downward-shifted copies of the equilibrium spectrum. The eigensystem of $M$ is therefore
\begin{align}
 M \vec{v}^{\alpha}_{j} = \left(\theta_{\alpha}-j \, z\right) \vec{v}^{\alpha}_{j} \, .
\end{align}
The eigenvectors $\vec{v}^{\alpha}_j$, are composed of $r+1$ vectors with the same dimension as $\bvec{\hat{g}}$ each. See \Eq{eq_eigenvector}.

The \RG flow is best interpreted when the parameters are written in terms of $\vec{v}^{\alpha}_j$. In particular, we can expand the vector denoting the microscopic distance from the fixed point in this basis
\begin{align}
 \vec{G}(\Lambda) = \vec{g}(\Lambda)-\vec{g}_{\text{fp}} = \sum_{\alpha j} \lambda_{\alpha j} \vec{v}^{\alpha}_j \, .
 \label{eq_projection}
\end{align}
$\Lambda$ is the scale at which the \RG flow is initiated and where the microscopic parameters are defined. This equation provides a definition of the parameters $\lambda_{\alpha j}$ as a linear combination of the microscopic parameters. In the critical region, the components of $\vec{G}(\Lambda)$ (and therefore $\lambda_{\alpha j}$) are small, and the flow is given by
\begin{align}
 \vec{G}(k) \cong \sum_{\alpha j} \lambda_{\alpha j} \, \left(\frac{k}{\Lambda}\right)^{\theta_\alpha-j z}\vec{v}^{\alpha}_j \, .
 \label{eq_flow} 
\end{align}
We have
\begin{align}
 \hat{\lambda}_{\alpha j}(k) = \lambda_{\alpha j} \left(\frac{k}{\Lambda}\right)^{\theta_\alpha-j z} \, .
\end{align}

${\lambda}_{\alpha j}$ and $\hat{\lambda}_{\alpha j}(k)$ are the parameters that we use in the main text for the $O(N)$ model, with the notation being
\begin{align}
 \tau_0 = \lambda_{10} \, , && \lambda_0 = \lambda_{20} \, , && \tau_1 = \lambda_{11} \, , && \lambda_1 = \lambda_{21} \, .
\end{align}
In particular, \Eq{eq_flow} becomes
\begin{align}
 \vec{G}(k) \cong \, & \tau_0 \left(\frac{k}{\Lambda}\right)^{-y} \hspace{-0.1cm}\vec{v}_0^1 + \lambda_0 \left(\frac{k}{\Lambda}\right)^{\varpi} \vec{v}_0^2 \nonumber \\
 & + \tau_1 \left(\frac{k}{\Lambda}\right)^{-y-z} \hspace{-0.1cm} \vec{v}_1^1 + c_2 \left(\frac{k}{\Lambda}\right)^{\varpi-z} \hspace{-0.1cm} \vec{v}_1^2 \, ,
 \label{eq_rg_flow}
\end{align}
which reproduces \Fig{fig_rg_flow}{} where $\hat{\lambda}_0$ is not represented. We see that $|\vec{G}(k)|$ increases as $k$ decreases and there is a scale $k_0$, below which \Eq{eq_rg_flow} is no longer applicable.  Away from the fixed point (and thus on spatial scales larger than $1/k_0$), the physics is nonuniversal. $k_0$ then separates the universal and nonuniversal regimes, and can be identified with the inverse correlation length ${\xi = 1/k_0}$. It can be estimated by picking the largest among the three values of $k$ for which the projections of $\vec{G}(k)$ along $\vec{v}_0^1$, $\vec{v}_1^1$, and $\vec{v}_1^2$ are, respectively $\pm 1$,
\begin{align*}
 1/\xi \sim k_0 = \Lambda \, \text{Max}\left[|\tau_0|^{1/y},\left|\tau_1\right|^{1/(y+z)},|\lambda_1|^{1/(z-\varpi)}\right] .
\end{align*}
This reproduces the three scaling regimes that are identified in the main text.

\subsection{Transversal and longitudinal drives}
\label{app_transverse_longitudinal}

We now provide some information on the eigenvectors of $M$ and the projection \Eq{eq_projection}. We will see that this provides the interpretation of the different scaling regimes as being transversal and longitudinal.

The eigenvectors of $M$ are expressed in terms of the eigenvectors of $M_0$ [see \Eq{eq_eigensystem}] and the off-diagonal elements of $M$. They take the following form
\begin{align}
 \vec{v}^{\alpha}_j = \left(\bvec{A}^{\alpha}_{j1},\bvec{A}^{\alpha}_{j2},\dots,\bvec{A}^{\alpha}_{jj-1},\bvec{v}^{\alpha},\bvec{0},\bvec{0},\dots \right) \, .
 \label{eq_eigenvector}
\end{align}
$\bvec{A}^{\alpha}_{jk}$ are vectors that depend on $\bvec{v}^{\alpha}$ and the different blocks of $M$. They can be computed recursively starting from $\bvec{A}^{\alpha}_{jj-1} = \left[\theta_{\alpha} - z - M_0\right]^{-1}M_{j-1,j}\bvec{v}^{\alpha}$. The important element of the above equation is that the $j^{\text{th}}$ sub-vector of $\vec{v}^{\alpha}_j$ is given by the equilibrium eigenvector $\bvec{v}^{\alpha}$.

From \Eq{eq_eigenvector} we find that the highest order part of the drive is a linear combination of the equilibrium eigenvectors. From \Eq{eq_projection} we can extract
\begin{align}
 \sum_{\alpha} \lambda_{{\alpha}r} \bvec{v}^{\alpha} = \bvec{\hat{g}}_r \, ,
 \label{eq_longitudinal_transverse}
\end{align}
because $\vec{g}_{\text{fp}} = (\bvec{g}_{\text{fp}},\bvec{0},\dots)$ only has nonvanishing components in its equilibrium ($i=0$) part. Moreover, the coordinates $\lambda_{{\alpha}r}$ do not depend on the lower-order drives $\bvec{\hat{g}}_{i<r}$. This is the main result of this appendix. It implies that the equilibrium eigenvectors provide the basis on which to decompose the $r^{\text{th}}$-order part of the drive. If $\bvec{g}_r$ is taken to be aligned enough with a given equilibrium eigenvector $\bvec{v}^{\alpha}$, then the nonequilibrium scaling will be $\xi \sim v^{-1/(r z-\theta_{\alpha})}$. The usual \KZ scaling emerges from a drive along the relevant direction and is therefore transversal. A drive that follows an irrelevant eigenvector will not cross the phase boundary and is longitudinal.


\end{document}